\documentclass[runningheads]{llncs}

\usepackage[T1]{fontenc}

\usepackage{graphicx}

\usepackage{url}

\usepackage{listings}
\usepackage{tcolorbox}
\usepackage{tikz}
\usepackage{booktabs}

\tcbuselibrary{skins}
\tcbuselibrary{breakable}

\begin{document}

\title{Unified Peripartum Database with Natural-Language-to-SQL Capabilities at Udine University Hospital: Design and Prototype}
\titlerunning{Unified Peripartum Database with Natural-Language-to-SQL Capabilities}

\author{Doriana Armenise\inst{2}
\and Ginevra Battello\inst{2}
\and Andrea Brunello\inst{1}
\and Lorenza Driul\inst{1,2}
\and Angelo Montanari\inst{1}
\and Elisa Rizzante\inst{2}
\and Nicola Saccomanno\inst{1}
\and Andrea Salvador\inst{1}
\and Serena Xodo\inst{2}
\and Silvia Zermano\inst{2}}

\authorrunning{A. Brunello \and N. Saccomanno \and A. Salvador}

\institute{Univeristy of Udine, 33100 Udine, Italy \and
Department of Maternal and Child Health, “Santa Maria della Misericordia” University Hospital, 33100 Udine, Italy}

\maketitle
\begin{abstract}

The fragmentation of obstetric information across electronic health record modules, device repositories, and laboratory systems, as it is common in hospitals, hinders both intrapartum care and reproducible research. In this work, we present a practical blueprint for transforming heterogeneous peripartum records into computable, queryable assets by designing and prototyping a unified peripartum relational database with natural-language-to-SQL (NL2SQL) capabilities at the Obstetrics Clinic of Udine University Hospital. Requirements were co-defined with clinicians and formalized as an Entity-Relationship diagram, from which the logical schema and SQL implementation of the database were then derived. The latter integrates heterogeneous sources to connect maternal anamnestic and longitudinal history, current-pregnancy findings, intrapartum course, and delivery and neonatal outcomes. The NL2SQL layer enables clinicians to pose natural-language queries to the system, lowering barriers to audit and exploratory 
analysis.

\end{abstract}

\section{Introduction}

The transition from traditional paper-based systems to electronic health records (EHRs) is reshaping obstetric and perinatal care, where the safety of both mother and newborn depends on a continuous, comprehensive flow of information. Far from being a mere digital filing cabinet, a EHR can pose as a foundational clinical asset for delivering safer, more efficient, and more effective care. 
A well-designed obstetric EHR consolidates fragmented data, including maternal history, antepartum risk factors, intrapartum monitoring (e.g., continuously recorded cardiotocography, CTG), and neonatal outcomes, into a single, accessible repository. This centralization enhances the continuity of care, ensuring that every clinician involved in the process has an immediate and complete understanding of the patient's status. For example, a complete clinical picture at a glance can aid in quick decision-making during a critical intrapartum event \cite{NICE2022NG229}. 
Beyond direct clinical care, EHRs provide an unprecedented wealth of longitudinal data for research and quality improvement. By aggregating standardized data across a large patient population, researchers can identify clinical trends, evaluate the effectiveness of interventions, and develop predictive models to anticipate and mitigate adverse events. This secondary use of data is a game-changer for advancing both clinical knowledge and patient safety.

However, despite their promise, implementing EHRs is complex and resource-intensive, with costs that extend beyond software licenses to hardware, infrastructure, integration, and sustained training and support. Equally important are sociotechnical challenges, including workflow redesign, change management, and usability. Disruptions to established practice can elicit resistance, and suboptimal interfaces or documentation burden increase cognitive load, erode situational awareness, and contribute to clinician frustration and burnout. Effective adoption thus requires co-design with end users, strong governance, and continuous optimization so that systems align with clinical workflows rather than the reverse. As a result of these difficulties, obstetric data often remain fragmented across heterogeneous modules, device archives, and laboratory systems, where coding variation, timestamp inconsistencies, and weak linkages hinder both quality audit, and research. Reviews and policy analyses underscore these issues and the need for maternity-specific interoperability guidance \cite{Glynn2019EHRHeterogeneity,Lewis2023EHRDQ,ONC2024MaternityInteroperability,OECD2023EHR,vanDerScheer2025PerinatalData}.

For these reasons, in this work, focusing on the use case of the Obstetrics Clinic at Udine University Hospital, we present the design and development of a centralized relational database for pregnancies and childbirths. The database has been co-designed with clinicians to ensure usability and data consistency across the peripartum domain, and it harmonizes data from heterogeneous sources, integrating $(i)$ maternal anamnestic and longitudinal data (e.g., pre-existing conditions, medications, obstetric/gynecologic history, prior pregnancy outcomes), $(ii)$ current-pregnancy data (e.g., laboratory results and visit outcomes), $(iii)$ intrapartum data (e.g., labor progress, CTG), and $(iv)$ delivery and neonatal outcomes (mode/timing of delivery, fetal blood cord gas analysis, neonatal status). To lower barriers at user (i.e., clinicians, obstetricians, and nurses) interaction, a natural-language-to-SQL (NL2SQL) layer lets medical personnel query the system with clinical prompts formulated in natural language, which are then automatically translated into Structured Query Language (SQL) queries. By providing a consistent and comprehensive view of obstetric data, the system supports analyses that can improve patient care by identifying patterns, trends, and risk factors, enabling the development of prevention tools \cite{corey2020using,mccullough2016health}, e.g., the early identification of fetuses at risk of acidemia, an area where precise operational thresholds remain limited and practice heterogeneous \cite{Cahill2018DecelArea,Chandraharan2007CTG,McCoy2025DeepLearningEFM}.

\section{Current state, limitations, and requirements}
\label{sec:udine}

\begin{figure}[t]
    \centering
\includegraphics[width=0.8\linewidth]{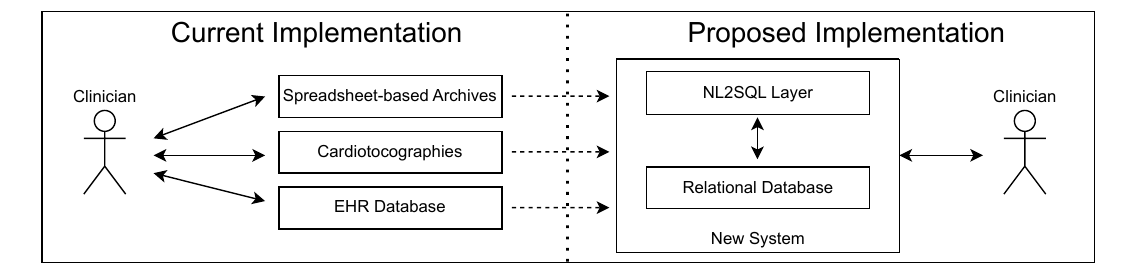}
    \caption{Current (left) and proposed (right) implementation of the information system. Solid arrows represent human-component or component-component interactions, while dashed arrows represent data flows.}
    \label{fig:implementation}
\end{figure}

Public hospitals in Italy rely on EHR modules that are typically integrated and standardized at least at the regional level (e.g., \cite{sesamo_fvg}). These systems facilitate interaction among healthcare professionals and, in some cases, patients themselves, while meeting legally mandated standards of security, performance, and accessibility. However, their data storage formats are primarily designed for day-to-day clinical operations rather than statistical or analytical purposes. Moreover, the way information is stored often differs from what clinicians find most practical for their workflows.

To compensate for these limitations, hospitals often implement internal data management solutions. Since at least 2022, clinicians at the Obstetrics Clinic of Udine University Hospital have complemented the official systems with manually compiled spreadsheets and custom tools to better organize data from check-ups and hospitalizations. While this approach facilitates quick access to frequently needed information, it has also resulted in multiple fragmented, ad hoc repositories. These informal archives, though adequate for routine care, lack the structured foundation required for advanced analyses and cross-patient comparisons, such as the use of unique identifiers, the enforcement of domain constraints, or the implementation of data models specifically designed to reduce redundancy.

As a result, any non-trivial task involves navigating a landscape of heterogeneous and non-integrated data sources, often redundant or inconsistent, a situation illustrated on the left side of Figure \ref{fig:implementation}. In the following, we present the various archives and modules currently in use within the Obstetrics Clinic, which serve as the foundation for our proposed integrated solution.

\smallskip
\noindent \textbf{Spreadsheets}
Clinicians rely on manually compiled, spreadsheet-based archives to record patient data during examinations. Collection of data typically occurs at specific stages of pregnancy: $(i)$ in the first trimester (approximately 11–13 weeks) for Nuchal Translucency (NT) screening, along with documenting pre-existing conditions and other related assessments; $(ii)$ in the second trimester (approximately 19–21 weeks) for morphological ultrasound and selected genetic screenings; and $(iii)$ during labor and delivery. Not all patients undergo each of these examinations at the Clinic, as some may choose other facilities within the hospital district. In addition, a subset of patients with high-risk pregnancies is monitored more frequently, with clinicians collecting detailed information on conditions, comorbidities, and ongoing therapies. As we previously mentioned, while spreadsheet-based archives provide basic functionalities and can be adapted to clinicians’ needs, their nature leads to significant challenges, including varying notations and interpretations across professionals, missing data, and conflicting entries arising from unchecked redundancies and constraints. 

\smallskip
\noindent \textbf{Cardiotocographies}
Delivery rooms are equipped with cardiotocographs \linebreak (CTGs) to monitor fetal and maternal heart rates as well as uterine contractions. CTGs can export data as spreadsheets containing measurements for each time point; in the delivery rooms of the Udine Obstetrics Clinic, these parameters are sampled at a frequency of four times per second. As we shall see, to ensure device independence, our database stores temporal data using a flexible, timestamp-based approach.

\smallskip
\noindent \textbf{EHR database}
Since 2022 clinicians have introduced a GUI-based EHR system, \emph{likely} supported by a relational database management system in the back end. However, it is not publicly distributed, and no documentation or references are available online. Data export is limited to spreadsheets, often inconsistent in format and style, and sometimes unreadable. Clinicians use this software only intermittently, as it lacks mechanisms for ensuring data consistency, particularly for chronologically ordered events, and allows redundancies. As a result, much of the information maintained in custom spreadsheets is never transferred into it. Ultimately, despite its ostensibly more structured design, this EHR database offers little improvement over spreadsheet-based archives regarding data integrity, constraint enforcement, or redundancy management, leaving clinicians with the same fundamental issues.

\smallskip 

Ultimately, the main challenges observed stem from fragmentation, internal and cross-system inconsistencies, and the absence of validation mechanisms in the current solutions. Users report difficulties in querying and cross-checking data, which hinders both the integration of individual patient histories and the development of advanced statistical analyses or predictive models. 
For example, patient and newborn identifiers are managed differently in spreadsheets and CTG tracings. Moreover, when data from spreadsheets is loaded into the EHR database, redundancies are introduced that can easily lead to inconsistencies.
Examples, populated with fictitious data, are provided in the Appendix.

The new information system must therefore enable clinicians to seamlessly store and retrieve all patient details currently captured in existing archives, while also incorporating additional elements identified through clinician interviews, such as biometric ultrasound data, extended neonatal physiological parameters, and a more comprehensive recording of pre-existing conditions beyond high-risk pregnancies. 
Furthermore, the system should be easily extensible to accommodate new data management requirements arising from clinical practice or research, and adaptable to evolving clinical needs; for example, by allowing clinicians to update the number and types of tests performed at each examination.
Finally, it must remain intuitive for clinicians to use in routine practice, as ease of use is paramount for successful adoption.

\section{Domain conceptual modeling}
\label{sec:conceptual}

\begin{figure}[t]
    \centering
    \includegraphics[width=1\linewidth]{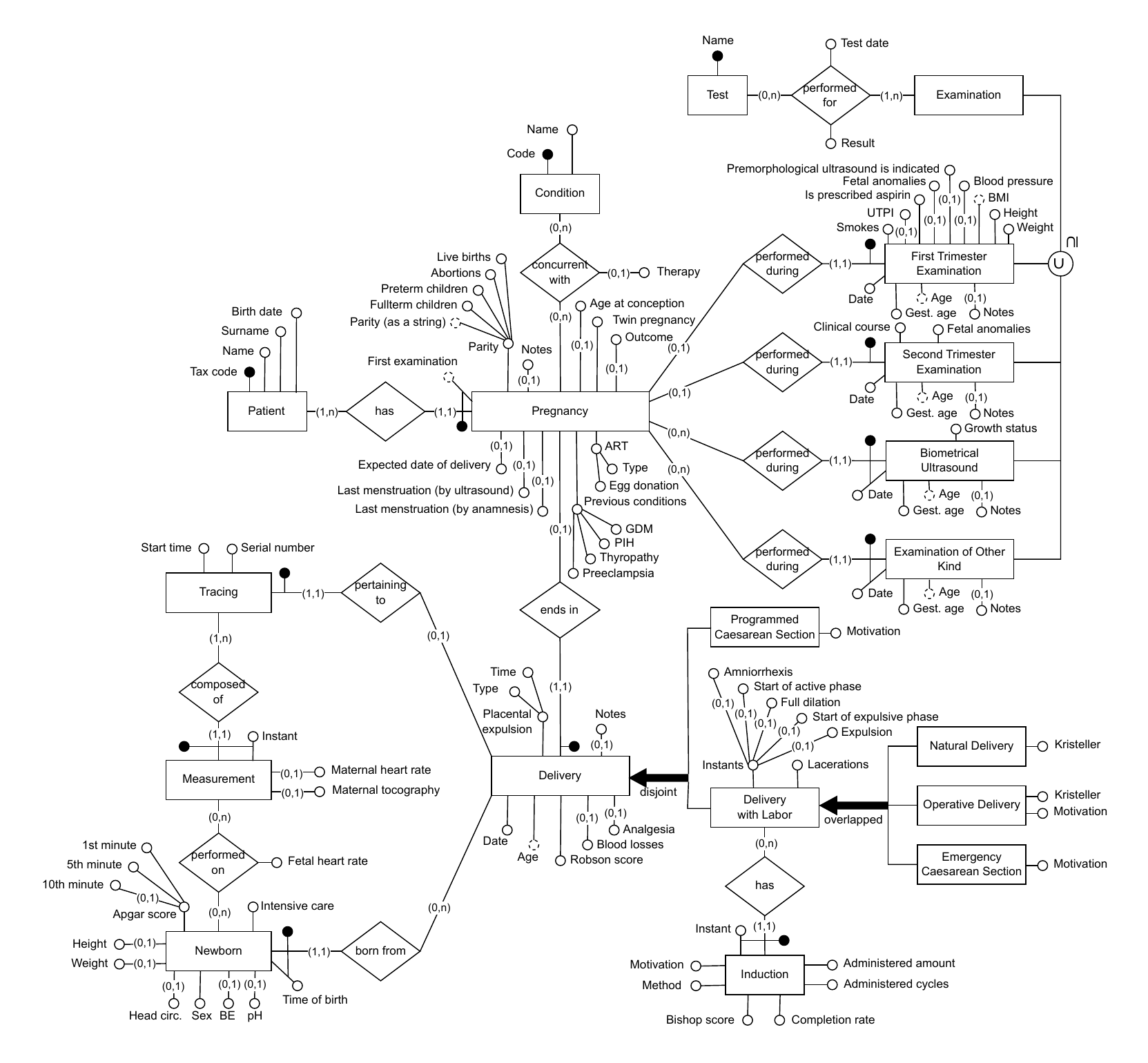}
    \caption{Entity-Relationship diagram of the analyzed domain.}
    \label{fig:erdiagram}
\end{figure}

This section presents the conceptual model developed in collaboration with clinicians, represented as an Entity–Relationship (ER) diagram and following the classical notation proposed in \cite{atzeni2000database}. The diagram and its documentation not only formally encode the domain requirements but also provide the foundation for the new centralized, integrated relational database. The diagram is shown in Figure~\ref{fig:erdiagram}, and described entity by entity in the following. 

\smallskip
\noindent \textbf{Patient}
The entity set \texttt{Patient} represents patients that have been examined at least once during pregnancy, or have given birth in the Udine Obstetrics Clinic.
Patients are identified by their Italian tax code, a unique 16-character alphanumeric string.\footnote{Of course, this identifier can be easily adapted for other contexts.} They are modeled with only a few essential anagraphic attributes, such as name, surname, and date of birth, as all other details are associated with related entities in the schema. As shown in Figure~\ref{fig:erdiagram}, a patient may experience one or more pregnancies, represented by the (1,n) relationship.

\smallskip
\noindent \textbf{Pregnancy}
The (weak) entity set \texttt{Pregnancy} represents pregnancies managed by the medical staff of the Udine Obstetrics Clinic.
Each pregnancy is associated with exactly one patient, as expressed by the (1,1) relationship in Figure~\ref{fig:erdiagram}. A pregnancy is uniquely identified by the patient and the date of the first examination, which is always recorded. 
Regarding the attributes of \texttt{Pregnancy}, the distinction between those pertaining to the overall pregnancy and those specific to examinations, particularly the nuchal translucency (NT) scan, can sometimes be ambiguous, since most pregnancy-related details are typically collected during the examinations.
In this work, we assigned to \texttt{Pregnancy} those attributes of a more general nature, while attributes that may vary over the course of pregnancy or have a context-specific meaning were assigned to the examination entities. This distinction is particularly important when, for instance, the NT examination is performed in another facility, as the “general” pregnancy data can still be recorded in the system. 
More in detail, \texttt{Pregnancy} includes information on parity,\footnote{Parity is the number of previous pregnancies, described as: $(i)$ number of full-term births, $(ii)$ number of premature births, $(iii)$ number of abortions, and $(iv)$ number of live births.} maternal age at conception, use of assisted reproductive technologies (if applicable), previous pregnancy-related conditions (e.g., preeclampsia 
or thyroid disease), date of last menstruation, and expected date of delivery.

\smallskip
\noindent \textbf{Condition}
The entity set \texttt{Condition} represents diseases or illnesses that may be present during pregnancy.
We associated \texttt{Condition} with the entity \texttt{Pregnancy}, rather than with \texttt{Patient}, since medical conditions may differ across successive pregnancies. This association is represented by the optional many-to-many relationship \texttt{concurrent with}. 
When a condition, present during a pregnancy, is under treatment, the corresponding therapy is recorded as an attribute of the relationship \texttt{concurrent with}.

\smallskip
\noindent \textbf{Examination}
The entity set \texttt{Examination} represents gynecological examinations or check-ups performed at the Clinic during pregnancy.
We identified four types of examinations: $(i)$ \texttt{First Trimester Examination}, primarily aimed at NT testing, performed at most once per pregnancy; $(ii)$ \texttt{Second Trimester \linebreak Examination}, focused on morphological ultrasound, also performed at most once; $(iii)$ \texttt{Biometrical ultrasound}, which may be performed multiple times during pregnancy; and $(iv)$ \texttt{Examination of Other Kind}.
These are modeled as a union-type entity \texttt{Examination} comprising the four examination types. This modeling approach is preferable to a generalization–specialization hierarchy, as it better accommodates the distinct cardinality constraints that characterize the relationships between the different examination types and the entity \texttt{Pregnancy}. 
For each entity of \texttt{Examination}, clinicians typically record the date, gestational age, and type-specific attributes.

\smallskip
\noindent \textbf{Test}
The entity set \texttt{Test} represents medical tests that can be performed during examinations. In particular, through the relationship \texttt{performed on}, one or more tests may be associated with an examination; conversely, a test may never have been performed (e.g., in the case of a test only recently introduced into clinical practice). The results of a performed test are modeled as attributes of the relationship \texttt{performed for}.

\smallskip
\noindent \textbf{Delivery}
The entity set \texttt{Delivery} represents childbirth events managed at the Clinic.
Its attributes depend on the delivery type. The first (total, disjoint) specialization distinguishes between \texttt{Programmed Caesarean Section} (performed before labor) and \texttt{Delivery with Labor}.
All deliveries include attributes such as gestational age, Robson score, placental expulsion, analgesia, and estimated blood loss. In addition, deliveries with labor record further attributes, including lacerations and key delivery time points.
A second (total, overlapping) specialization applies to deliveries with labor, distinguishing \texttt{Natural Delivery} (spontaneous vaginal births), \texttt{Operative Delivery} (assisted vaginal births), and \texttt{Emergency Caesarean Section}. The overalap between these types is possible, for example, in twin pregnancies where different procedures are required for each neonate. Each subtype has its own specific attributes.

\smallskip
\noindent \textbf{Induction}
The entity set \texttt{Induction} represents labor inductions, which are absent in patients undergoing scheduled Caesarean sections.
It records details such as timing (partial key), method, drug dosage, and completion rate.

\smallskip
\noindent \textbf{Newborn}
The entity set \texttt{Newborn} represents a child delivered.
In multiple pregnancies (e.g., twins or triplets), newborns are distinguished by their time of birth.
Both physical (e.g., weight, length) and physiological (e.g., Apgar score, blood pH) data are recorded here.

\smallskip
\noindent \textbf{Tracing}
The entity set \texttt{Tracing} represents a complete cardiotocography (CTG) recorded during delivery. Machine-dependent information, such as the specific fetal sensors used, is not modeled, as it constitutes technical data that has not been considered relevant from the domain perspective. Conceptually, a \texttt{Tracing} serves as a container for measurements.

\smallskip
\noindent \textbf{Measurement}
The entity set \texttt{Measurement} represents the individual data points that compose a CTG, each associated with its corresponding timestamp.
Attributes stored directly in \texttt{Measurement} include maternal heart rate and uterine contractions, whereas fetal heart rates are modeled as attributes of the relationship \texttt{performed on}, so to link them to the specific \texttt{Newborn} they refer to. 
Only observed values are taken into account; missing values are not stored.

\smallskip
The weak entities \texttt{Newborn}, \texttt{Delivery}, \texttt{Tracing}, and \texttt{Measurement} form a cycle; to avoid inconsistencies, when relationship are instanced, they must all refer to the same delivery.
As we will see, when translated into a logical schema, foreign key constraint are sufficient in guaranteeing this correspondence.

\section{Development of the new system}
\label{sec:relational}

In this section, we present the development of the new system, which comprises two core components (Figure~\ref{fig:implementation}, right): $(i)$ a relational database, designed according to the developed Entity–Relationship diagram, capable of integrating information from all previous sources; and $(ii)$ an NL2SQL layer that facilitates clinician interaction with the database.
All code related to the system implementation, as well as a prototypical deployment of the database, is available at: \url{https://github.com/dslab-uniud/gineco-db}.

\subsection{Relational database}

From the ER diagram, a logical relational schema was derived. 
Then, the schema was implemented in a PostgreSQL~\cite{postgresql} instance.

\begin{figure}[t]
    \centering
    \includegraphics[width=1\linewidth]{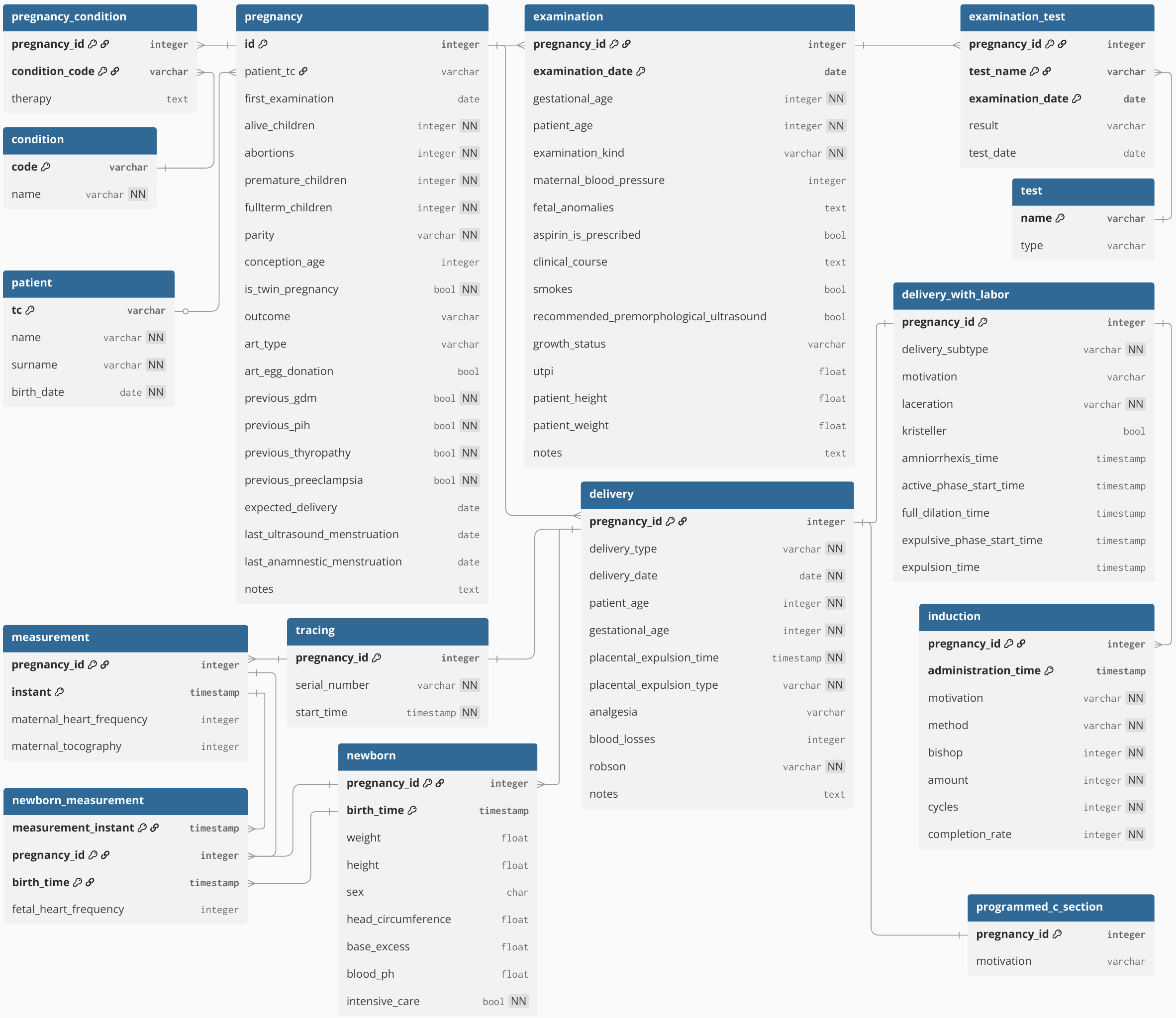}
    \caption{Logical relational diagram of the analyzed domain.}
    \label{fig:relationaldiagram}
\end{figure}

\smallskip
\noindent \textbf{Logical schema}
The schema, shown in Figure~\ref{fig:relationaldiagram}, was derived primarily by following standard mapping rules \cite{atzeni2000database}. It adopts the well-known Crow's Foot notation \cite{silberschatz2010db}. 
The main design choices made in defining the schema are:
$(i)$ we assigned a synthetic \texttt{id} to each pregnancy, rather than relying on the natural ER identifier. This choice reduces the number of attributes in related tables, since many of them include the pregnancy identifier as a foreign key;
$(ii)$ given the similarity of their attributes, the four types of examinations were consolidated into a single relation, thereby simplifying the logical schema;
$(iii)$ since tests are modeled as a separate entity, the attribute \texttt{result} must be able to assume many data types; the attribute \texttt{type} in \texttt{Test} dictates the actual data type of the outcome\footnote{
We define three categories of tests according to their outcome, which may be numeric, string, or enumerated. Since the physical data type of the attribute \texttt{result} is \texttt{varchar} (string), numeric values are cast to \texttt{varchar} as well.
} and triggers check the correspondence between the two.
$(iv)$ since multiple induction methods may be applied in a single delivery, they were modeled in a separate relation;
$(v)$ and, the two-layer specialization of deliveries was simplified into three relations: all deliveries, containing the common attributes; programmed Caesarean sections, representing cases without labor; and deliveries with labor, which may include labor inductions.

As previously mentioned, the cycle of foreign keys formed between \texttt{Newborn}, \texttt{Delivery}, and \texttt{Tracing} requires no additional checks, as the foreign key constraints alone ensure that tuples pertain to the same delivery.

Not all domain requirements can be enforced directly by the logical schema. To ensure data consistency, we defined additional checks in the final database implementation, using triggers. They can be found in the Appendix.

\smallskip
\noindent \textbf{Physical implementation}
We implemented our proposed database using PostgreSQL \cite{postgresql} database management system. 
Table definitions in SQL were derived directly from the logical schema presented in Section \ref{sec:relational}.
An interesting aspect concerns the handling of the attribute \texttt{type} in the \texttt{test} table. This attribute, implemented as an array of \texttt{varchar}, determines the domain of the attribute \texttt{result} in the \texttt{examination\_test} table. Specifically, if \texttt{type} is set to \texttt{[string]}, results can take any \texttt{varchar} value; if it is set to \texttt{[numeric]}, \texttt{result} must be a string interpretable as a numeric value; in all other cases, \texttt{type} enumerates the exact list of values that \texttt{result} may assume. 
For the complete physical implementation of the database, as well as a working prototype, we refer the reader to the previously mentioned online resource.

\subsection{NL2SQL module}

The database schema was developed in close collaboration with clinicians to ensure that it meets their practical needs. However, interacting with a database typically requires writing SQL code, a task that can be non-trivial for users without sufficient training in database querying. To address this challenge, the new system integrates an NL2SQL module, which accepts queries expressed in natural language, translates them into SQL, executes them on the database, and returns the results as a table. The module consists of two main components: a front-end graphical user interface (GUI) that facilitates user interaction, and a back-end deep learning model that performs the translation.

\smallskip
\noindent \textbf{Front-end GUI}
We chose SQLChat \cite{sqlchat2025} as the front-end. SQLChat is a highly configurable web-based application: it connects to relational databases (e.g., PostgreSQL, MySQL) and to AI models exposed via any OpenAI-compatible API endpoint, also supporting self-hosted and locally run models, paramount requirements for privacy-sensitive contexts, like ours. The interface relies on the familiar chat paradigm common to LLM agents. 
For each session, SQLChat automatically establishes a context that describes the schema of the active database and includes directives guiding the model’s responses, ensuring that user queries are answered in a database-aware manner. For further details, see the Appendix.

\smallskip
\noindent \textbf{Back-end model}
The model chosen as the NL2SQL back-end is XiYanSQL-QwenCoder-32B-2504, an open-source SQL generation model with multi-dialect support and high performance on common SQL test datasets;  specifically, the authors self-report a performance of 67.14\% on BIRD Dev compared to, e.g., 60.37\% obtained by OmniSQL-32B or 59.39\% with GPT-4.1-0414.
This model is presented as part of XiYanSQL \cite{XiYanSQL}, a wider NL2SQL framework which provides finetuning capabilities, but it is also distributed as a stand-alone component.

\section{Usage of the new system}
\label{sec:proto}

A prototype of the newly developed system is already been tested by the Obstetrics Clinic personnel. Data from the original sources were migrated into the system using Python scripts, which can also be executed periodically to update and synchronize its content. This is intended as a temporary solution until the new system fully replaces the legacy ones.
In the following, we present a representative example of successful interaction supported by the system, while a more extensive discussion is provided in the Appendix.

 The query reports the number of C-sections performed in 2024, which may result either from a programmed delivery or from an urgent intervention during labor. The model employs an \texttt{EXISTS} operator to verify whether a given C-section is associated with a delivery that occurred in 2024. In fact, each C-section is linked to exactly one delivery, so a \texttt{JOIN} operation would have sufficed. This constraint is enforced in our system through a trigger (see Appendix) rather than explicitly represented in the schema, and is therefore unavailable to the NL2SQL model. In future developments, we plan to mitigate that by enriching the database schema provided in the prompt template to the model with comments that capture such additional information.
\begin{tcolorbox}[title=Query: \texttt{C-sections in 2024},
  beforeafter skip=5pt,
  boxsep=0.5mm,
  fontupper=\small,
  colback=gray!5!white,
  colframe=gray!75!black,
  enhanced,
  breakable]

Count the number of C-sections, both programmed and with labor, performed in 2024.

\tcbsubtitle{SQL query automatic translation}
\begin{verbatim}
SELECT COUNT(*) AS total_c_sections 
FROM (SELECT 1 
      FROM programmed_c_section AS pcs
      WHERE EXISTS (SELECT 1 
                    FROM delivery AS d
                    WHERE pcs.pregnancy_id = d.pregnancy_id 
                        AND EXTRACT(YEAR FROM d.delivery_date) = 2024)    
    UNION ALL
      SELECT 1 
      FROM delivery_with_labor AS dwl
      WHERE EXISTS (SELECT 1 
                    FROM delivery AS d
                    WHERE dwl.pregnancy_id = d.pregnancy_id 
                        AND EXTRACT(YEAR FROM d.delivery_date) = 2024
                        AND d.delivery_type = 'emergency_c_section' ) 
    ) AS c_sections;
\end{verbatim}
\end{tcolorbox}

\smallskip 

Overall, out of eight attempted moderate-complexity queries, the system failed to answer correctly only one, provided that the user formulates an unambiguous and comprehensive natural-language request explicitly referring to the relevant table and attribute names. 
Of course, beyond the interaction capabilities offered by the NL2SQL module, advanced users can still submit custom SQL queries. In addition, parameterized stored procedures for recurring queries have already been integrated into the database.

\section{Related work}
\label{sec:rel}

In the literature, works have addressed data management and decision support in obstetrics and gynecology, but with emphases different from ours. 
Public datasets have been proposed to support research in maternal health (e.g., \cite{chudavcek2014open}). While valuable for statistical analysis and predictive modeling, these resources are not designed as operational databases for clinical use. 
Existing proposals for obstetrics and gynecology information systems are typically modeled around the patient as a general medical entity rather than specifically around pregnancy and newborns \cite{skarga2015modeling,vida2012modeling}, while others focus on nursing workflows \cite{zhu2020design}. Some contributions describe high-level architectures for hospital information systems in obstetrics~\cite{barrote2014obstetric}, while earlier studies evaluated perinatal data collection systems \cite{maresh1986assessment}. However, these contributions do not provide a pregnancy-centered database design. 
Since the late 1980s, expert systems have also been explored to support medical reasoning in obstetrics and gynecology \cite{lam2015architecture,riss1988development}. These approaches concentrate primarily on clinical decision support, often relying on rule-based or ontology-driven reasoning, rather than on the development of a full-fledged information system. 
In fact, recent reviews emphasize the continued need for decision-support systems in obstetrics and gynecology, stressing both human-centered design \cite{cockburn2024clinical} and the importance of clinical implementation and the integration of historical data \cite{lin2024artificial}.

Aligned with these observations, and differing from previous contributions, our work proposes a blueprint for an information system centered on pregnancy, explicitly considering clinical usability 
and laying the foundation for future computer-assisted, possibly AI-based, decision support.

\section{Conclusions}

In this work, we presented the design and development of a centralized relational database for pregnancies and childbirths at Udine University Hospital's Obstetrics Clinic. The system integrates heterogeneous maternal, intrapartum, and neonatal data into a consistent structure, co-designed with clinicians to ensure usability. By integrating an NL2SQL interface, the novel system lowers barriers for clinical users and enables both routine practice support and research applications. This lays the foundation for improved risk assessment, decision support, and predictive modeling in obstetrics, with the potential to enhance maternal and neonatal outcomes. A prototypical implementation of the database is available online. In the future, this implementation could evolve into a hub for the continual publication of properly anonymized datasets, fostering reproducible research and enabling the broader scientific community to develop and validate novel methods for perinatal care.

\begin{credits}

\bigskip

\noindent \textbf{Ethics declaration.}
This study has been approved by the Institutional Review Board of the University of Udine's Department of Medicine (Prot. IRB: 122/2025).

\smallskip

\noindent \textbf{\ackname} AB, AM, NS acknowledge the support from the Interconnected Nord-Est Innovation Ecosystem (iNEST),
    which received funding from the European Union Next-GenerationEU (PIANO NAZIONALE DI RIPRESA E RESILIENZA 
    – MISSIONE 4 COMPONENTE 2,
    INVESTIMENTO 1.5 – D.D. 1058 23/06/2022, ECS00000043).
    AM acknowledges the support from the MUR PNRR project FAIR - Future AI Research
    (PE00000013) also funded by the European Union Next-GenerationEU.  This manuscript reflects only the authors’ views and opinions, neither the European Union nor the European Commission can be considered responsible for them.

\smallskip

\noindent \textbf{\discintname}
The authors have no relevant competing interests to declare.
\end{credits}

\bibliographystyle{splncs04}
\bibliography{mybibliography.bib}

\begin{thebibliography}{10}
\providecommand{\url}[1]{\texttt{#1}}
\providecommand{\urlprefix}{URL }
\providecommand{\doi}[1]{https://doi.org/#1}

\bibitem{atzeni2000database}
Atzeni, P., Ceri, S., Paraboschi, S., et~al.: Database Systems: Concepts, Languages and Architectures. McGraw-Hill (2000), \url{https://dbbook.inf.uniroma3.it/}

\bibitem{barrote2014obstetric}
Barrote, A., Silva, P., Gon{\c{c}}alves, F., et~al.: Obstetric information system: {E}ffectiveness in health care practice. Procedia Technology  \textbf{16},  1411--1416 (2014)

\bibitem{Cahill2018DecelArea}
Cahill, A.G., Tuuli, M.G., Stout, M.J., et~al.: A prospective cohort study of fetal heart rate monitoring: {D}eceleration area is predictive of fetal acidemia. American Journal of Obstetrics and Gynecology  \textbf{218}(5),  523.e1--523.e12 (2018)

\bibitem{Chandraharan2007CTG}
Chandraharan, E., Arulkumaran, S.: Prevention of birth asphyxia: {R}esponding appropriately to cardiotocograph ({CTG}) traces. Best Practice \& Research Clinical Obstetrics \& Gynaecology  \textbf{21}(4),  609--624 (2007)

\bibitem{sqlchat2025}
Chen, T., et~al.: {SQLChat}: {C}hat-based {SQL} client and editor for the next decade. \url{https://github.com/sqlchat/sqlchat} (2025), accessed: 2025-08-27

\bibitem{chudavcek2014open}
Chud{\'a}{\v{c}}ek, V., Spilka, J., Bur{\v{s}}a, M., et~al.: Open access intrapartum {CTG} database. BMC Pregnancy and Childbirth  \textbf{14}(1), ~16 (2014)

\bibitem{cockburn2024clinical}
Cockburn, N., Osborne, C., et~al.: Clinical decision support systems for maternity care: {A} systematic review and meta-analysis. eClinicalMedicine  \textbf{76} (2024)

\bibitem{corey2020using}
Corey, L., Vezina, A., Gala, R.B.: Using technology to improve women's health care. Ochsner Journal  \textbf{20}(4),  422--425 (2020)

\bibitem{Glynn2019EHRHeterogeneity}
Glynn, E.F., Hoffman, M.A.: Heterogeneity introduced by {EHR} system implementation in a de-identified data resource from 100 non-affiliated organizations. JAMIA Open  \textbf{2}(4),  554--561 (2019)

\bibitem{lam2015architecture}
Lam, J., Abdullah, M.S., Supriyanto, E.: Architecture for clinical decision support system {(CDSS)} using high risk pregnancy ontology. ARPN Journal of Engineering and Applied Sciences  \textbf{10}(3),  1229--1237 (2015)

\bibitem{Lewis2023EHRDQ}
Lewis, A.E., et~al.: Electronic health record data quality assessment and tools: {A} systematic review. Journal of the American Medical Informatics Association  \textbf{30}(10),  1730--1740 (2023)

\bibitem{lin2024artificial}
Lin, X., Liang, C., Liu, J., et~al.: Artificial intelligence-augmented clinical decision support systems for pregnancy care: {S}ystematic review. Journal of medical Internet research  \textbf{26},  e54737 (2024)

\bibitem{XiYanSQL}
Liu, Y., Zhu, Y., Gao, Y., et~al.: {XiYan-SQL}: {A} novel multi-generator framework for text-to-{SQL} (2025), \url{https://arxiv.org/abs/2507.04701}

\bibitem{maresh1986assessment}
Maresh, M., Dawson, A., Beard, R.: Assessment of an on-line computerized perinatal data collection and information system. BJOG: An International Journal of Obstetrics \& Gynaecology  \textbf{93}(12),  1239--1245 (1986)

\bibitem{McCoy2025DeepLearningEFM}
McCoy, J.A., Levine, L.D., Wan, G., et~al.: Intrapartum electronic fetal heart rate monitoring to predict acidemia at birth with the use of deep learning. American Journal of Obstetrics and Gynecology  \textbf{232},  116.e1--116.e9 (2025)

\bibitem{mccullough2016health}
McCullough, J.S., Parente, S.T., Town, R.: Health information technology and patient outcomes: the role of information and labor coordination. The RAND Journal of Economics  \textbf{47}(1),  207--236 (2016)

\bibitem{NICE2022NG229}
{National Institute for Health and Care Excellence}: Fetal monitoring in labour. \url{https://www.nice.org.uk/guidance/ng229} (2022), accessed: 2025-08-12

\bibitem{ONC2024MaternityInteroperability}
{Office of the National Coordinator for Health IT (ONC)}: Interoperability of maternity health care records: {B}est practices informational resource. \url{https://www.healthit.gov/sites/default/files/page/2024-03/FY23%20ONC%20Maternity%20Care%20Info%20Resource%20Guide_508%20(1).pdf} (2024)

\bibitem{postgresql}
{PostgreSQL Global Development Group}: {PostgreSQL} documentation. \url{https://www.postgresql.org/docs/current/} (2025), accessed: 2025-08-24

\bibitem{sesamo_fvg}
{Regione Autonoma Friuli Venezia Giulia}: {SESAMO}, {E}lectronic {P}latform for {H}ealth {S}ervices. \url{https://sesamo.sanita.fvg.it/sesamo/} (2018), accessed: 2025-08-20

\bibitem{riss1988development}
Riss, P.A., Koelbl, H., Reinthaller, A., et~al.: Development and application of simple expert systems in obstetrics and gynecology. Journal of Perinatal Medicine  (1988)

\bibitem{silberschatz2010db}
Silberschatz, A., Korth, H.F., Sudarshan, S.: Database System Concepts. McGraw-Hill, New York, 6 edn. (2010)

\bibitem{skarga2015modeling}
Skarga-Bandurova, I., Biloborodova, T., Nesterov, M.: Modeling structures for integrated obstetrics, gynecology, and neonatal information system. Problemy informatsiinykh tekhnologii  \textbf{17}(1),  51--57 (2015)

\bibitem{OECD2023EHR}
Slawomirski, L., et~al.: Progress on implementing and using electronic health record systems: {D}evelopments in {OECD} countries as of 2021. Tech. rep., OECD Health Working Papers No. 160 (2023)

\bibitem{vanDerScheer2025PerinatalData}
Van Der~Scheer, J.W., Komolafe, V., Webster, K., et~al.: Improving {UK} data on avoidable perinatal brain injury: {R}eview of data dictionaries and consultation. Pediatric Research  (2025)

\bibitem{vida2012modeling}
Vida, M., Mioara Stoicu-Tivadar, L., Blobel, B., et~al.: Modeling the framework for obstetrics-gynecology department information systems. European Journal for Biomedical Informatics  \textbf{8}(3),  57--64 (2012)

\bibitem{zhu2020design}
Zhu, H., Gu, L.: Design and application of intelligent information system for comprehensive management of obstetrics and gynecology health care. Journal of Medical Imaging and Health Informatics  \textbf{10}(8),  1834--1840 (2020)

\end{thebibliography}

\clearpage
\newpage

\appendix
\section*{Appendix}

Here we present the supplementary material of the article. Please note that all developed code pertaining to the new system implementation, as well as a working prototype of the database, can be found at:
\url{https://github.com/dslab-uniud/gineco-db}.

\section{Details on the systems in use by the clinicians}

Here we present some extracts from spreadsheets currently in use at the Clinic, as discussed in Section~\ref{sec:udine}. The aim is twofold: to illustrate the specific information clinicians consider essential, and to highlight the lack of structure in the existing systems that our proposal seeks to address. For the sake of clarity, the text has been translated into English; however, note that the original files contain a mixture of English and Italian labels, which further reduces their overall readability.

\bigskip

Table~\ref{tab:firsttrimester} presents a snippet from the \emph{first trimester} spreadsheet, which is intended to record anamnestic and examination data from the early stage of pregnancy. Here, missing values are inconsistently represented, appearing either as double slashes or as zeros.

\bigskip

Table~\ref{tab:delivery} shows a snippet from the delivery spreadsheet. This example illustrates data duplication, as both episiotomy and analgesia are represented by two columns, one storing a Boolean marker (presence/absence) and another containing a string value. In addition, the column \texttt{Apgar score} may include either two or three measurements, depending on whether an assessment was also performed at the tenth minute.

\bigskip

Table~\ref{tab:ehrbase1} presents a CSV excerpt extracted from the EHR database. The attribute \texttt{Outcome}, which is conceptually categorical, appears as integer values that represent internal codes without any explicit mapping to human-readable categories. Moreover, the attributes from \texttt{PIH} to \texttt{Thyropathy}, although Boolean at a conceptual level, are encoded inconsistently: they are left empty when false, while a seemingly arbitrary integer is inserted when true. In fact, this integer corresponds to an internal identifier assigned to the pregnancy.

\bigskip

Finally, Table~\ref{tab:ehrbase2} presents another excerpt from the EHR database. Many rows exhibit misaligned values, with entries belonging to the \texttt{Newborn section} frequently appearing outside their corresponding column. In addition, missing data are inconsistently represented, as they are sometimes explicitly marked as \texttt{missing}, and other times simply left blank.

\begin{table}[!ht]
    \centering
    \caption{Example snippet from the first trimester spreadsheet.}
    \begin{tabular}{|l|l|l|l|}
        \hline
         \textbf{Outcome of genetic tests} & \textbf{Premorph. ultr. indicated?} & \textbf{NIPT} & \textbf{Outcome} \\
         \hline
         // & 0 & 1 & B \\
         // & 0 & 0 & // \\
         // & 0 & 1 & B \\
         amniocentesis: normal karyotype & 0 & 0 & // \\
         // & 0 & 1 & B \\
         // & 1 because of FHR < 5°P & 0 & // \\
         \hline
    \end{tabular}
    \label{tab:firsttrimester}
\end{table}

\begin{table}[!ht]
    \centering
    \vspace{1em}
    \caption{Example snippet from the delivery spreadsheet.}
    \begin{tabular}{|l|l|l|l|l|}
        \hline
         \textbf{Episiotomy} & \textbf{Motivation} & \textbf{Analgesia} & \textbf{Type} & \textbf{Apgar score}  \\
         \hline
         1 & rigid perineum & NO & & 9-9 \\
         0 & & NO & & 9-10 \\
         0 & & YES & epidural & 9-10 \\
         0 & & YES & epidural & 9-10 \\
         & & YES & epidural & 6-8-9 \\
         \hline
    \end{tabular}
    \label{tab:delivery}
\end{table}

\begin{table}[!ht]
    \centering
    \vspace{1em}
    \caption{First example snippet from EHR database, exported as a CSV file.}
    \begin{tabular}{|l|l|l|l|l|}
        \hline
         \textbf{Outcome} & \textbf{PIH} & \textbf{ART} & \textbf{GDM} & \textbf{Thyropathy}  \\
         \hline
         6 & & & 612 & \\
         4 & 616 & & & \\
         4 & & & & 713 \\
         6 & & 859 & 859 & \\
         \hline
    \end{tabular}
    \label{tab:ehrbase1}
\end{table}

\begin{table}[!ht]
    \centering
    \vspace{1em}
    \caption{Second example snippet from EHR database, exported as a CSV file.}
    \begin{tabular}{|l|l|l|l|l|}
        \hline
         \textbf{Newborn Section} & \textbf{Apgar 1} & \textbf{Apgar 5} & \textbf{Apgar 10} & \textbf{Weight gr}  \\
         \hline
         74 & 5 & Newborn Section & 9 & 10 \\
         73 & 73 & 5 & Newborn Section & 9 \\
         & & 2 & Newborn Section & 8 \\
         Newborn Section & 8 & 9 & missing & 3300 \\
         28 & 28 & 2 & Newborn Section & 7 \\
         27 & 18 & Newborn Section & 9 & 9 \\
         \hline
    \end{tabular}
    \label{tab:ehrbase2}
\end{table}

\section{Further constraints implemented in our developed database}

Not all domain constraints are enforced directly by the logical schema. To ensure data consistency, we defined and implemented additional checks in the final database through the use of triggers. The corresponding code is available in the online system reference; here, we provide an intuitive description:
\begin{itemize}
    \item For each instance of \texttt{pregnancy} there must exist at least one corresponding instance in \texttt{examination} or a corresponding instance in \texttt{delivery}.
    \item For each instance of \texttt{pregnancy} there must exist at most one corresponding first trimester examination (i.e., an instance of \texttt{examination} where the field \texttt{examination\_kind} is \texttt{first\_trimester}). The same applies for the second trimester examination.
    \item For each instance of \texttt{examination\_test} the data type of \texttt{result} is coherent with the data type contained in the corrisponding instance in \texttt{test}.
    \item For each instance of \texttt{delivery} there exists exactly one instance of either \texttt{programmed\_c\_section} or \texttt{delivery\_with\_labor}.
    \item For each instance of \texttt{measurement} there exists at least one recorded value (which may be in the attributes \texttt{maternal\_heart\_rate}, \linebreak \texttt{maternal\_tocography}, or in an instance of \texttt{newborn\_measurement}).
\end{itemize}

\section{SQLChat front-end}

The NL2SQL module invokes the XiYanSQL-QwenCoder model via the SQLChat interface. Upon connecting to a database, SQLChat constructs a context prompt that includes the database schema and passes it to the model. 

\begin{figure}[t]
    \centering
    \includegraphics[width=0.9\linewidth]{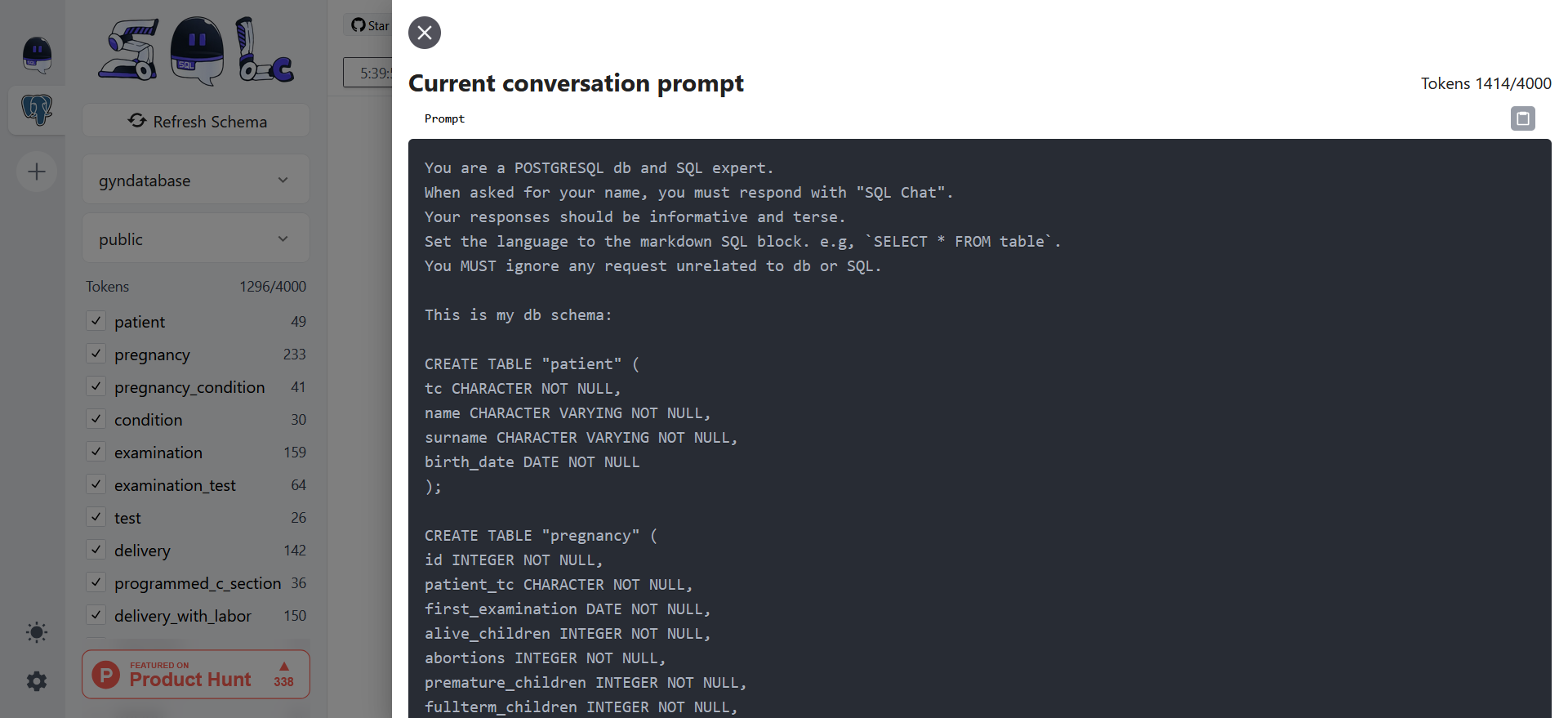}
    \caption{Screenshot of the SQLChat GUI. The panel \texttt{Current conversation prompt} shows the beginning of the automatically generated context prompt.}
    \label{fig:sqlchat1}
\end{figure}

\begin{figure}[!ht]
    \centering
    \includegraphics[width=0.9\linewidth]{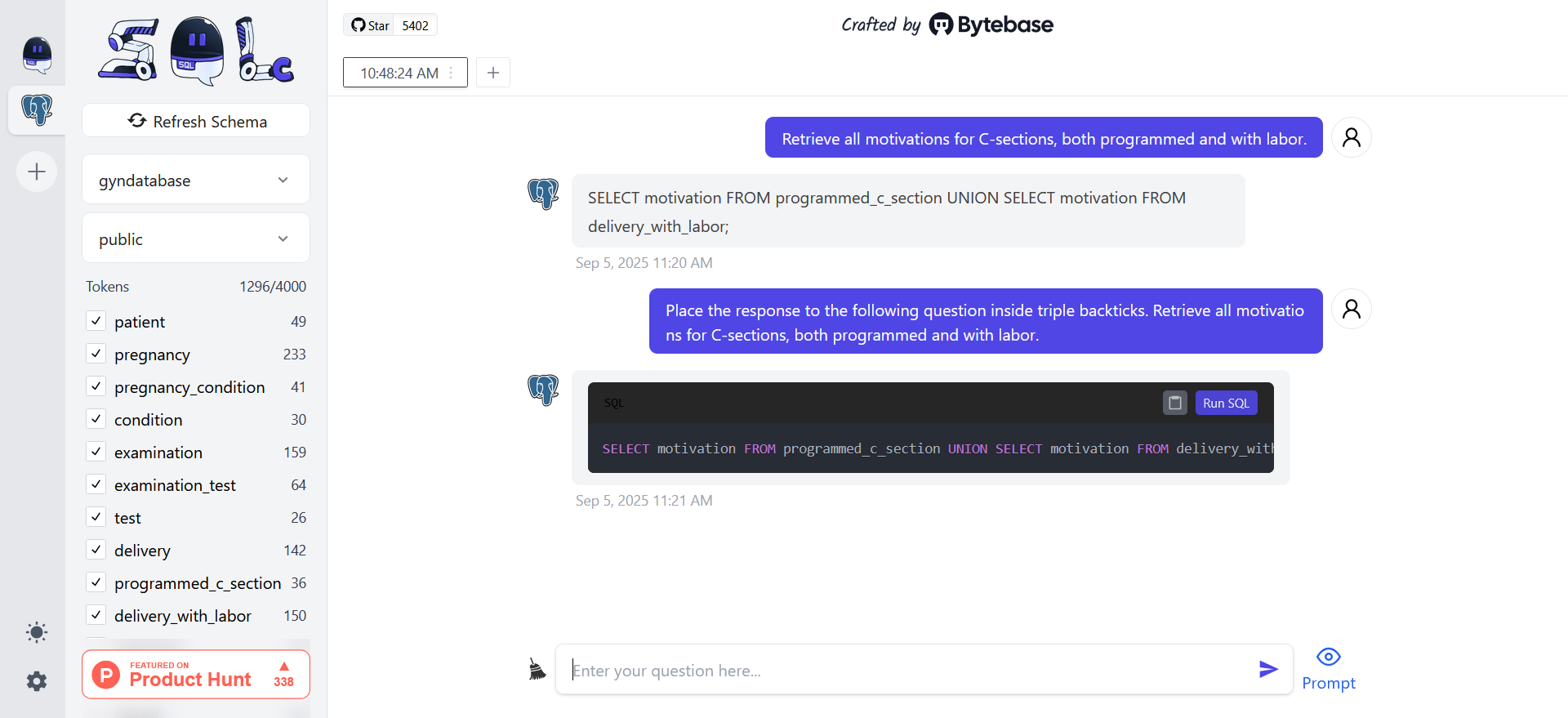}
    \caption{Screenshot of the SQLChat GUI. The main panel shows the dialogue between the user and the NL2SQL agent.}
    \label{fig:sqlchat2}
\end{figure}

\begin{figure}[!ht]
    \centering
    \includegraphics[width=0.9\linewidth]{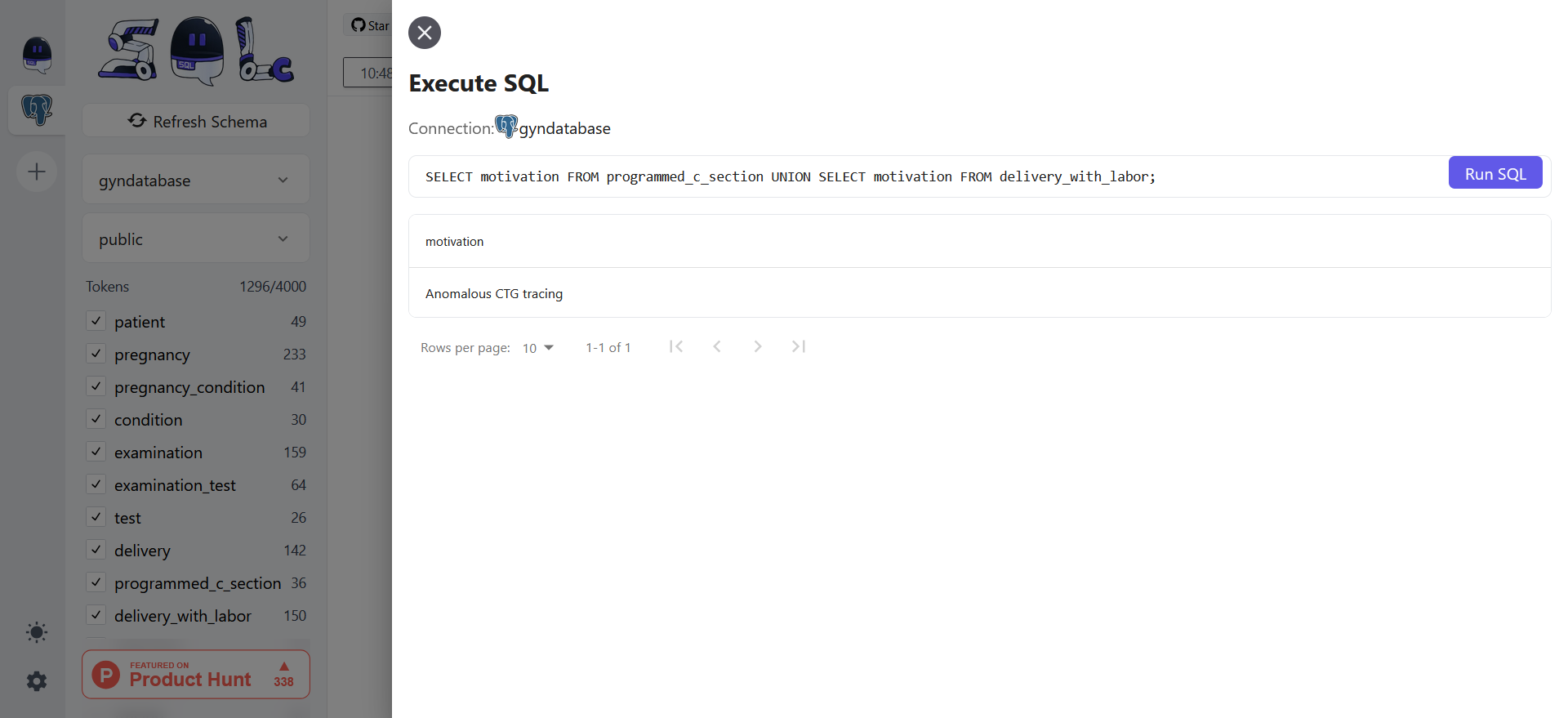}
    \caption{Screenshot of the SQLChat GUI. The panel \texttt{Execute SQL} shows query text and result after execution.}
    \label{fig:sqlchat3}
\end{figure}

The schema is provided as standard SQL Data Definition Language (DDL) statements, derived directly from the logical schema in Section~\ref{sec:relational} (see Figure~\ref{fig:sqlchat1}). User queries are then issued to the model through a chat-style dialogue (see Figure~\ref{fig:sqlchat2}).
Agent responses formatted as SQL queries can be executed against the database and result is shown in a separate panel (see Figure~\ref{fig:sqlchat3}); generated SQL queries can be edited manually.

\section{Extended system interactions}

In addition to the query presented in the main article, we discuss here further examples of meaningful interactions enabled by the system.

The first query retrieves all deliveries in which the newborn’s blood pH is lower than 7.1, a value likely indicative of acidemia.
\begin{tcolorbox}[title=Query: \texttt{anomalous blood pH after delivery},
  beforeafter skip=5pt,
  boxsep=0.5mm,
  fontupper=\small,
  colback=gray!5!white,
  colframe=gray!75!black,
  enhanced,
  breakable]

List the name of the patients and the date of delivery for all deliveries in which the newborn has a pH lower than 7.1

\tcbsubtitle{SQL query automatic translation}
\begin{verbatim}
SELECT p.name, d.delivery_date
FROM delivery d 
    JOIN newborn n ON d.pregnancy_id = n.pregnancy_id
    JOIN patient p ON d.pregnancy_id = p.tc
WHERE n.ph < 7.1;
\end{verbatim}

\end{tcolorbox}

\bigskip
For C-sections and operative deliveries, it is important to understand the underlying reasons for which they were performed. The following query should focus on C-sections. Note how, nevertheless, the query is imprecise as, for deliveries with labor it also retrieves motivations for operative deliveries.
\begin{tcolorbox}[title=Query: \texttt{C-sections motivations},
  beforeafter skip=5pt,
  boxsep=0.5mm,
  fontupper=\small,
  colback=gray!5!white,
  colframe=gray!75!black,
  enhanced,
  breakable]

Retrieve all motivations for C-sections, both programmed and with labor.

\tcbsubtitle{SQL query automatic translation [incorrect]}
\begin{verbatim}
    SELECT motivation 
    FROM programmed_c_section 
UNION
    SELECT motivation 
    FROM delivery_with_labor;
\end{verbatim}

\tcbsubtitle{SQL query manually corrected translation}
\begin{verbatim}
    SELECT motivation 
    FROM programmed_c_section 
UNION
    SELECT motivation 
    FROM delivery_with_labor
    WHERE delivery_subtype = 'emergency_c_section';
\end{verbatim}

\end{tcolorbox}

\bigskip
During delivery, lacerations may occur. Clinicians are interested in understanding how many first-, second-, third-, and fourth-degree lacerations, as well as how many episiotomies and trachelorrhaphies, occur across the total number of vaginal deliveries, operative deliveries, and emergency C-sections performed after failed vacuum extraction. The following query focuses on lacerations.
\begin{tcolorbox}[title=Query: \texttt{Number of lacerations},
  beforeafter skip=5pt,
  boxsep=0.5mm,
  fontupper=\small,
  colback=gray!5!white,
  colframe=gray!75!black,
  enhanced,
  breakable]

Count the number of lacerations and their percentage relative to deliveries with labor.

\tcbsubtitle{SQL query automatic translation [correct]}
\begin{verbatim}
SELECT laceration, 
       COUNT(*), 
       ROUND((COUNT(*) * 100.0 / (SELECT COUNT(*) 
                                  FROM delivery_with_labor)), 2)
FROM delivery_with_labor
GROUP BY laceration
ORDER BY count DESC;
\end{verbatim}
\end{tcolorbox}

\bigskip
The following three queries focus on inductions performed during delivery. Specifically, the first retrieves the absolute and relative number of inductions with respect to the total number of deliveries. The second computes the average duration of the interval between induction administration and expulsion. Finally, the third reports the number of inductions undergone by each patient who delivered in 2025.
\begin{tcolorbox}[title=Query: \texttt{Number of deliveries with inductions},
  beforeafter skip=5pt,
  boxsep=0.5mm,
  fontupper=\small,
  colback=gray!5!white,
  colframe=gray!75!black,
  enhanced,
  breakable]

Count the number of deliveries that had inductions and their percentage with respect to all deliveries.

\tcbsubtitle{SQL query automatic translation [correct]}
\begin{verbatim}
SELECT COUNT(DISTINCT i.pregnancy_id) AS induced_deliveries_count,
       ROUND((COUNT(DISTINCT i.pregnancy_id) * 100.0
              / COUNT(DISTINCT d.pregnancy_id)), 2)
              AS induced_deliveries_percentage
FROM induction i 
        JOIN delivery d ON i.pregnancy_id = d.pregnancy_id;
\end{verbatim}
\end{tcolorbox}

\begin{tcolorbox}[title=Query: \texttt{Average duration of induction-expulsion intervals},
  beforeafter skip=5pt,
  boxsep=0.5mm,
  fontupper=\small,
  colback=gray!5!white,
  colframe=gray!75!black,
  enhanced,
  breakable]

Calculate the average length of the interval between induction administration time and delivery expulsion time.

\tcbsubtitle{SQL query automatic translation [correct]}
\begin{verbatim}
SELECT AVG(EXTRACT(
           EPOCH FROM (d.expulsion_time - i.administration_time)))
           / 3600 AS average_interval_hours
FROM induction i 
        JOIN delivery_with_labor d ON i.pregnancy_id = d.pregnancy_id;
\end{verbatim}
\end{tcolorbox}

\begin{tcolorbox}[title=Query: \texttt{Number of inductions per patient in 2025},
  beforeafter skip=5pt,
  boxsep=0.5mm,
  fontupper=\small,
  colback=gray!5!white,
  colframe=gray!75!black,
  enhanced,
  breakable]

Calculate the number of inductions undergone by each patient who delivered in 2025.

\tcbsubtitle{SQL query automatic translation [correct]}
\begin{verbatim}
SELECT p.tc AS patient_tc, 
       p.name AS patient_name, 
       p.surname AS patient_surname, 
       COUNT(i.pregnancy_id) AS number_of_inductions 
FROM patient p
        JOIN delivery d ON p.tc = (SELECT patient_tc                               
                                   FROM pregnancy                            
                                   WHERE id = d.pregnancy_id) 
        JOIN induction i ON d.pregnancy_id = i.pregnancy_id
WHERE EXTRACT(YEAR FROM d.delivery_date) = 2025
GROUP BY p.tc, p.name, p.surname 
ORDER BY p.tc;
\end{verbatim}
\end{tcolorbox}

\bigskip
Finally, for clinical purposes, it is important to investigate whether there are commonalities among patients who underwent operative deliveries or C-sections for CTG-related reasons. To such an extent, information about them and their pregnancies can be extracted.
\begin{tcolorbox}[title=Query: \texttt{Pregnancy data of patients with CTG-related delivery complications},
  beforeafter skip=5pt,
  boxsep=0.5mm,
  fontupper=\small,
  colback=gray!5!white,
  colframe=gray!75!black,
  enhanced,
  breakable]

Show patient data and pregnancy data for all patients whose delivery motivation, whether C-section or operative, mentions CTG in any manner.

\tcbsubtitle{SQL query automatic translation [correct]}
\begin{verbatim}
SELECT p.*, 
       pr.* 
FROM patient p
    JOIN pregnancy pr ON p.tc = pr.patient_tc 
    JOIN delivery d ON pr.id = d.pregnancy_id 
    LEFT JOIN programmed_c_section pcs
                ON d.pregnancy_id = pcs.pregnancy_id 
    LEFT JOIN delivery_with_labor dwl
                ON d.pregnancy_id = dwl.pregnancy_id 
WHERE pcs.motivation ILIKE '%CTG%' OR dwl.motivation ILIKE '%CTG%';
\end{verbatim}
\end{tcolorbox}

\end{document}